\documentclass[article,showpacs,10pt]{revtex4}
\usepackage{amssymb}
\usepackage{epsfig}
\usepackage{graphicx}

\usepackage{bm}

\begin{document}

\title[Understanding the edge effect in TASEP with mean-field theoretic approaches]{Understanding the edge effect in TASEP with mean-field theoretic approaches}

\author{J J Dong$^{1,2}$, R K P Zia$^2$ and B Schmittmann$^2$}
\address{$^1$Department of Physics, Hamline University, St. Paul, MN,
55104, USA.}
\address{$^2$Department of Physics, Virginia Tech, Blacksburg, VA 24061,
USA.}

\email{jdong01@hamline.edu}

\begin{abstract}
We study a totally asymmetric simple exclusion process (TASEP) with one
defect site, hopping rate $q<1$, near the system boundary. Regarding our
system as a pair of uniform TASEP's coupled through the defect, we study
various methods to match a \emph{finite} TASEP and an \emph{infinite} one
across a common boundary. Several approximation schemes are investigated.
Utilizing the finite segment mean-field (FSMF) method, we set up a framework
for computing the steady state current $J$ as a function of the entry rate $%
\alpha $ and $q$. For the case where the defect is located at the entry
site, we obtain an analytical expression for $J\left( \alpha ,q\right) $
which is in good agreement with Monte Carlo simulation results. When the
defect is located deeper in the bulk, we refined the scheme of MacDonald,
et.al. [Biopolymers, \textbf{6}, 1 (1968)] and find reasonably good fits to
the density profiles before the defect site. We discuss the strengths and
limitations of each method, as well as possible avenues for further studies.
\end{abstract}

\vspace{2pc}
\maketitle
\section{Introduction}

Since its inception nearly four decades ago, the totally asymmetric simple
exclusion process (TASEP) \cite{Spitzer,Derrida92,DEHP,S1993,Derrida,Schutz}
has become a paradigmatic model in non-equilibrium statistical mechanics.
Not only is it one of the few mathematically tractable models in this field,
it displays a rich variety of behaviors and provides insight to a range
of complex physical systems, e.g., interface growth \cite{KPZ,WolfTang},
biopolymerization \cite{MG,LBSZia,TomChou,ShawMF,LBS} and traffic \cite
{Chowdhury,Popkov}. In its simplest form, TASEP consists of particles
hopping uni-directionally and stochastically on a one-dimensional (1D)
lattice with complete exclusion (each site accommodating no more than a
single particle). The original model \cite{Spitzer} was defined on a ring
(periodic 1D lattice) and, despite having a trivial steady state
distribution, displays complex \emph{dynamical} phenomena. In a TASEP with
open boundaries, there are even richer phenomena. Coupled to an infinite
reservoir, particles enter/leave the lattice with rate $\alpha $/$\beta $
(relative to the hopping rate within the lattice). The stationary state
distribution was found analytically through a matrix ansatz \cite{DEHP} and
displays three distinct phases along with continuous and discontinuous
transitions \cite{Schutz}.

Independent of Spitzer \cite{Spitzer}, a more general version of the open
TASEP was proposed \cite{MG} to model the translation process in protein
synthesis. In a living cell, the genetic code in the DNA is transcribed into
messenger RNA's (mRNA's), which are then used to synthesize proteins
(strings of amino acids) by a process which closely resembles a TASEP.
However, to model this biological process properly, at least two major
generalizations are required. Referring the reader to existing literature 
\cite{MG,LBSZia,TomChou,ShawMF,LBS,DSZ} for the details, we only state these
differences here: \\(i) Each particle ``covers'' $\ell >1$ sites (typically
12 \cite{MG,Heinrich,Kang}), i.e., exclusion occurs at a distance $\ell $.\\%
(ii) The hopping rates are inhomogeneous, i.e., the rate for a particle at
site $i$ to hop (provided site $i+\ell $ is empty) is $\gamma _i$ and is
expected to depend on the codon at $i$.\\Even the seemingly simple
modification in (i) is so serious that an exact steady state distribution
remains illusive. Only Monte Carlo simulations and mean field theories
provide good estimates of certain steady state properties \cite
{LBSZia,Lakatos}. Though many properties are qualitatively similar to the $%
\ell =1$ case, such as displaying three phases (maximal current, MC, and
low/high density, LD/HD) in the thermodynamic limit, there are important
quantitative differences. For example, the phase boundaries in the $\alpha $-%
$\beta $ phase diagram shift to 
\begin{equation}
\hat{\chi}\equiv \frac 1{1+\sqrt{\ell }}  \label{chi-hat}
\end{equation}
i.e., MC prevails for $\alpha ,\beta >\hat{\chi}$, LD for $\alpha <\min
(\beta ,\hat{\chi})$, HD for $\beta <\min (\alpha ,\hat{\chi})$. The average
overall density $\rho $ and current $J$ are also modified. In MC, we have $%
\rho =1-\hat{\chi}$ and $J=\hat{\chi}^2$. In LD, $\rho $ is now $\alpha \ell
/(1+\alpha \bar{\ell})$ , where 
\begin{equation}
\bar{\ell}\equiv \ell -1.  \label{l-bar}
\end{equation}
For HD, $\rho $ actually remains the same: $1-\beta $. In all cases, the
current is given by 
\begin{equation}
J(\rho )=\rho (1-\rho )/(\ell -\bar{\ell}\rho ).  \label{infJ}
\end{equation}
Beyond these simple quantities, the profiles are affected by $\ell >1$ quite
seriously \cite{ShawMF,DSZ,DSZ_PRE}.

Clearly, generalization (ii) is much more intractable. A further
complication is that the genetic code is not a ``random'' sequence. Thus, it
is unclear if the notion of quenched random averages \cite{HS,ShawMF,Foul} -
so successful in the studies of spin glasses \cite{glass} - is even
meaningful here. Nevertheless, from the point of view of physics, it is
reasonable to ask what the effects of inhomogeneities are on an open TASEP
with extended particles. Along these lines, there have been several studies
using different methods, on a variety of systems. Examples include a single
``defect'' in an otherwise uniform TASEP \cite{Kolo,TomChou,LBS,DSZ}, two
defects \cite{TomChou,DSZ}, a cluster of defects \cite{Lakatos,GS}, as well
as a fully inhomogeneous set $\left\{ \gamma _i\right\} $ that is dictated
by real genetic sequences \cite{LBSZia,ShawMF,DSZ}. In this context, we
re-examine the open TASEP with a single defect here, i.e., $\gamma _k\equiv
q\neq 1;\gamma _{i\neq k}=1$. In particular, the steady state current is
naturally suppressed if $q<1$, but further, simulation studies found that
this suppression is not as severe when the defect is located near the
entrance (typically $k\lesssim O(10)$) or the exit. This phenomenon was
coined the ``edge effect''\cite{DSZ}.

In this article, we focus on understanding this effect better. Since exact
solutions are not available, we consider several levels of approximations,
providing increasingly accurate predictions for the currents and density
profiles. We also discuss the effects of introducing $\ell >1$ particles
into the system. The remainder of this paper is organized as follows. In the
next section, we provide some details of our model and a brief summary of
previous results. As our approximations consist of neglecting certain
correlations, we regard them as different levels of ``mean field theories.''
Two new levels are presented in Section \ref{MCMF}. In Section \ref{sum}, we
end with a summary and outlook for future research.

\section{\label{model}The model and simulation details}

Our model consists of a 1D lattice of $N$ sites, with open boundaries. Each
site, labeled by $i=1,2,...,N$, is either occupied or vacant, so that a
configuration is specified by the familiar set of occupation numbers $%
\left\{ n_i\right\} ,\,n_i=1,0$. However, unlike the standard lattice gas
model, we have particles of size $\ell $, in the sense that a single
particle always occupies (or ``covers'') $\ell $ consecutive sites.
Therefore, strong correlations in $\left\{ n_i\right\} $ necessarily appear;
not all $2^N$ possibilities of $\left\{ n_i\right\} $ are allowed. A further
complication is that, in the most-often used specification \cite
{LBSZia,Lakatos}, a particle must lie fully on the lattice on the left
(i.e., occupying $i=1,...,\ell $) but it can ``dangle beyond'' the right
(i.e., only the particle's left most site must lie within the lattice). As a
result, the total number of holes on the lattice can vary even for a given,
fixed number of particles. This complication can be ameliorated, however, if
we choose the lattice to have $N+\bar{\ell}$ sites and symmetrize the rules
for entrance and exit. To conform with the notation of previous studies, we
will avoid this route here.

An alternative specification is to locate each particle by one of the $\ell $
sites, e.g., its left most site. With protein synthesis in mind, we follow 
\cite{LBSZia} and refer to this special site (on the particle) as the
``reader.'' The motivation comes from the ribosome ``reading'' the next
codon (and waiting for the arrival of the associated transfer RNA) before it
can move onto the next codon. With this convention, we define $r_i=1$ if
site $i$ is occupied by a reader and $r_i=0$ otherwise. Clearly, $\left\{
r_i\right\} $ labels a configuration and, like $\left\{ n_i\right\} $, there
are strong correlations. Choosing to locate the reader at the left end of a
particle \cite{LBSZia}, $\left\{ n_i\right\} $ can be generated from $%
\left\{ r_i\right\} $ by $n_{i+j}=r_i$ for $j\in \left[ 0,\ell -1\right] $.
We will also use the reader position to locate the particle, so that $r_i=1$
will be used interchangeably with ``A particle is located at site $i$.''
Finally, we define all sites beyond $N$ to be free, so that a particle at
the last $\ell $ sites is not hindered sterically by any others.

Turning to the dynamic rules, it is easiest to state them in terms of $r_i$.
The motion of an interior particle is obvious; only the entry/exit rules
need clarification. Coined ``complete entry, incremental exit'' in Ref. \cite
{Lakatos}, these are: \\

\begin{itemize}
\item  $r_1\rightarrow 1$ with rate $\alpha $, \emph{provided} $r_k=0,$ for $%
k\in \left[ 1,\ell \right] $;

\item  $r_i=1\rightarrow r_{i+1}=1$ with rate $\gamma _i$, \emph{provided} $%
r_{i+\ell }=0,$ for $i\in \left[ 1,N-\bar{\ell}-1\right] $;

\item  $r_i=1\rightarrow r_{i+1}=1$ with rate $\gamma _i$, for $i\in \left[
N-\bar{\ell},N-1\right] $;

\item  $r_N=1\rightarrow r_N=0$ with rate $\beta $.
\end{itemize}

\begin{figure}[tbp]
\begin{center}
\includegraphics[height=3cm,width=10cm]{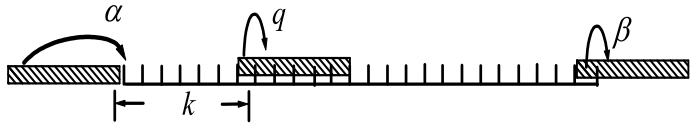} \hspace{-2cm}
\end{center}
\caption{Sketch of a TASEP for particle size $\ell =6$ with a single slow
site at position $k$, with rate $q$.}
\label{fig:one-slow}
\end{figure}

In our simulations, we establish an array of $N$ entries to represent the
lattice sites, as well as an extra one ($i=0$) for the reservoir. We use a
random sequential updating scheme and keep track of the locations of
readers. In one Monte Carlo step (MCS), we make $M+1$ attempts to update,
where $M$ is the total number of particles on the lattice. As the $1$
accounts for a particle in the reservoir to be chosen, there is an even
chance for each particle to be updated once, as well as introducing a new
particle into the system. A sketch of this process is shown in Fig.~\ref
{fig:one-slow}. The lattice is initially empty and we discard the first $%
2\times 10^6$ MCS to ensure that the system has reached the steady state. A
further $2\times 10^6$ MCS are used for collecting measurements, each
separated by $100$ MCS in order to avoid temporal correlations. Unless
otherwise noted, averaging over the $2\times 10^4$ measurements provides
good statistics. Such steady state averages will be denoted by $\left\langle
...\right\rangle $. We studied different system sizes between $200$ and $%
1000 $, with most data taken from $N=1000$. 

To characterize the state of the system, we monitor several observables. The
most obvious is 
\begin{equation}
\rho _i^{\text{r}}\equiv \left\langle r_i\right\rangle ,  \label{rho r}
\end{equation}
a quantity we will refer to as the reader density. Of course, $\sum_i\rho
_i^{\text{r}}$ is just the average number of particles in the system (i.e.,
ribosomes on the mRNA). Thus, the overall particle density $\frac
1N\sum_i\rho _i^{\text{r}}$ has an upper bound of $1/\ell $. Another
interesting variable, $\rho _i\equiv \left\langle n_i\right\rangle $,
referred to as the ``coverage density'', is the probability that site $i$ is
covered by a particle (regardless of the location of the reader). Of course,
it carries the same information as $\rho _i^{\text{r}}$, since the two are
related through 
\begin{equation}
\left\{ 
\begin{array}{l}
\rho _i=\sum_{k=0}^{\ell -1}\rho _{i-k}^{\text{r}} \\ 
\rho _i^{\text{r}}=\rho _i-\rho _{i-1}+\rho _{i-\ell }^{\text{r}}
\end{array}
\right.  \label{cr}
\end{equation}
(with the understanding $\rho _i^{\text{r}}\equiv 0$ for $i\leq 0$). The 
\emph{overall} coverage density, $\rho \equiv \frac 1N\sum_i\rho _i$, may reach unity
and provides a good indication of how packed the system is. From $\rho _i$,
we can also access the profile for the vacancies (holes): 
\begin{equation}
\rho _i^{\text{h}}=1-\rho _i\,\,.  \label{rho h}
\end{equation}

When $\ell =1$, the two density profiles are of course identical. As soon as 
$\ell >1$, serious correlations appear \cite{ShawMF,DSZ,DSZ_PRE}.

A quantity of great importance to a biological system is the steady state
level of a given protein. If we assume that the degradation rates are
(approximately) constant under certain growth condition, then these levels
are directly related to the protein production rates. In our model, such a
rate is just the average particle current $J$, defined as the average number
of particles exiting the system per unit time. At steady state, it is also
the current measured across any section of the lattice. For simplicity and
to ensure the best statistics, we count the total number of particles which
enter the lattice over the entire measurement period.

For our investigations here, we focus on one simple type of inhomogeneity: a
single ``slow'' site (Fig.~\ref{fig:one-slow}) in an otherwise homogeneous
lattice, i.e., a bottleneck along a smooth road. Locating the defect at site 
$k$, we have 
\begin{equation}
\gamma _{i\neq k}=1\quad \text{and}\quad \gamma _k=q  \label{gammas here}
\end{equation}
with $q<1$ . A common approach to this type of problems is to study the
lattice as two sublattices (left, sites $1$ to $k$, denoted by $L$, and
right, the rest, denoted by $R$) connected by $q$ and having the same
through current $J_L=J_R$. We are especially interested in the dependence of
the current, denoted by $J(q,k)$, on the parameters $q$ and $k$.

Previous studies located the defect far from the system boundaries, e.g., $%
k\approx N/2$ \cite{Kolo,LBS,DSZ_PRE}. There, it is sufficient to regard
both sublattices as infinite and to exploit the results of the single TASEP
while matching $L$ and $R$ appropriately. Of course, the matching condition
is not exactly known and the previous studies propose different
approximation schemes. These approaches lead to tolerably good predictions
for the average densities and currents. Here, we provide examples for the $%
\alpha =\beta =1$ case. In the most naive scheme (referred to as the ``naive
mean-field,'' NMF, approximation in \cite{DSZ_PRE}), the exact expression
for the current, $q\left\langle r_k\left( 1-n_{k+\ell }\right) \right\rangle 
$, is replaced by $q\left\langle r_k\right\rangle \left\langle 1-n_{k+\ell
}\right\rangle $. Despite the severity of this approximation, the result for
the current 
\begin{equation}
J_{\text{NMF}}=\left\{ 
\begin{array}{cc}
q/\left[ (1+q)(1+q\ell )\right] & \text{ for }q\leq 1/\sqrt{\ell } \\ 
\hat{\chi}^2 & \text{ for }q\geq 1/\sqrt{\ell }
\end{array}
\right. \,  \label{J_NMF}
\end{equation}
captures all the features of the system qualitatively. An alternative
approach \cite{LBS} takes into account some of the correlations in $%
\left\langle r_kn_{k+\ell }\right\rangle $, leading to 
\begin{equation}
J_{\text{SKL}}=\left. \left( 1-Q-\sqrt{1-2Q}\right) \right/ \left( Q\bar{\ell%
}\right)  \label{J_SKL}
\end{equation}
where $Q\equiv \left. 2q\bar{\ell}\left( 1+q\bar{\ell}\right) \right/
\,\left( 1+q+2q\bar{\ell}\right) ^2$. The agreement with data is much better
than NMF, as shown explicitly in \cite{DSZ_PRE}.

However, it is clear that neither scheme can provide any information on how
the system is affected by the \emph{location }of the defect, $k$. Now,
simulations with $\ell $ up to $12$ showed a non-negligible increase \cite
{DSZ,DSZ_PRE,GS} in $J$ as the defect approaches the system boundaries, a
phenomenon coined the ``edge effect'' in \cite{DSZ}. In the next section, we
consider two more refined approaches. One is based on a mean-field theory
proposed by MacDonald, Gibbs and Pipkin (MGP) \cite{MG}, which we
generalized to an inhomogeneous TASEP. The other, proposed by Chou \cite
{TomChouprivate}, consists of a cluster approximation (FSMF, finite-segment
mean-field \cite{TomChou}), which accounts for the ``interaction'' between
the entry rate $\alpha $ and the defect $q$. Though both provide much
improvement over the expression in Eqn.(\ref{J_NMF}) above, their
limitations will be discussed. 

\section{\label{MCMF}Beyond simple mean field theory}

To appreciate the various levels of approximations, we begin with the exact
expressions for the current. From the dynamic rules, $J$ is given generally
by 
\begin{eqnarray}
J &=&\alpha \left\langle 1-n_\ell \right\rangle =\alpha \rho _\ell ^{\text{h}%
}  \label{exactJ1} \\
&=&\gamma _i\left\langle r_i\left( 1-n_{i+\ell }\right) \right\rangle
\,\,;\quad i\in \left[ 1,N-\ell \right]  \label{exactJ2} \\
&=&\gamma _i\left\langle r_i\right\rangle \,\,;\quad \quad i\in \left[ N-%
\bar{\ell},N-1\right]  \label{exactJ3} \\
&=&\beta \left\langle r_N\right\rangle \,\,.  \label{exactJ4}
\end{eqnarray}
If we define $\gamma _N\equiv \beta $, then the last two equations are just $%
J=\gamma _i\rho _i^{\text{r}}$ for the last $\ell $ sites.

In the study here, we have $\gamma _{i\neq k}=1$ and small $k$. So, Eqn. (%
\ref{exactJ2}) becomes 
\begin{eqnarray}
J &=&\left\langle r_i\left( 1-n_{i+\ell }\right) \right\rangle \,\,;\quad
i\in \left[ 1,N-\ell \right] ,i\neq k  \label{exactJnek} \\
&=&q\left\langle r_k\left( 1-n_{k+\ell }\right) \right\rangle \,.\,
\label{exactJq}
\end{eqnarray}

\subsection{\label{sub:rr}An approach using the MGP recursion relation}

In the naive mean field approach, $\left\langle r_in_j\right\rangle $ is
replaced by $\left\langle r_i\right\rangle \left\langle n_j\right\rangle $.
For $\ell =1$, this approximation turns out to be quite good. Thus, the
profile is well described by the solution to the one-term recursion
relation, namely $J_{\text{NMF}}=\rho _i\left( 1-\rho _{i+1}\right) $.
Unfortunately, this naive approach fails for $\ell >1$. The difficulty is
due in part to $\left\langle r_i\right\rangle \neq \left\langle
n_i\right\rangle $ and in part to the severe exclusion at $\ell >1$. MGP
took into account some of this exclusion \cite{MG} and proposed a much
better approximation. The key lies in replacing $p\left( \text{ }h_k\text{ }|%
\text{ }r_k\,\text{\text{ or }}\,h_k\right) $, the conditional probability
of finding a hole at site $k$ \textit{given} that this site is occupied by 
\emph{either a reader or a hole}, according to the fraction: 
\begin{equation}
p\left( h_k\text{ }|r_k\,\text{\text{ or }}\,h_k\right) \cong \frac{\rho _i^{%
\text{h}}}{\rho _i^{\text{r}}+\rho _i^{\text{h}}}\,\,.  \label{p rh}
\end{equation}
Notice that, for $\ell =1$, the denominator is simply unity and an equality
holds. Now, $\left\langle r_i(1-n_{i+\ell })\right\rangle $ is given by the
product of $\rho _i^{\text{r}}$ and $p\left( h_{i+\ell }|r_i\right) $, the
conditional probability of having a hole at site $i+\ell $ \textit{given}
there is a reader at site $i$. But, if a reader exists at site $i$, then
site $i+\ell $ must be occupied by \emph{either a hole or a reader}, so that 
$p\left( h_{i+\ell }|r_i\right) =p\left( h_{i+\ell }\text{ }|r_{i+\ell }\,%
\text{\text{ or }}\,h_{i+\ell }\right) $. With the key approximation above,
MGP's scheme can be summarized by 
\begin{equation}
\left\langle r_i(1-n_{i+\ell })\right\rangle =\rho _i^{\text{r}}p\left(
h_{i+\ell }|r_i\right) \rightarrow \frac{\rho _i^{\text{r}}\rho _{i+\ell }^{%
\text{h}}}{\rho _{i+\ell }^{\text{r}}+\rho _{i+\ell }^{\text{h}}}\,\,.
\label{MG_approx}
\end{equation}
Generalizing to the inhomogeneous case, an approximate Eqn. (\ref{exactJ2})
now reads: 
\begin{equation}
J_{\text{MGP}}=\gamma _i\frac{\rho _i^{\text{r}}\rho _{i+\ell }^{\text{h}}}{%
\rho _{i+\ell }^{\text{r}}+\rho _{i+\ell }^{\text{h}}}\,\,.  \label{MG_rec}
\end{equation}

To proceed, we follow MGP and regard this as an $\ell $-term recursion
relation. Starting from the last site, Eqns. (\ref{exactJ3},\ref{exactJ4})
allow us to write down the first $\ell $ terms: 
\begin{equation}
\rho _i^{\text{r}}=J_{\text{MGP}}/\gamma _i\,;\quad \quad i\in \left[ N-\bar{%
\ell},N\right] \,\,.  \label{start}
\end{equation}
With these, we can start a ''backwards'' recursion (BR) relation: 
\begin{equation}
\rho _i^{\text{r}}=\frac{J_{\text{MGP}}}{\gamma _i}\left\{ 1+\frac{\rho
_{i+\ell }^{\text{r}}}{1-\sum_{k=1}^\ell \rho _{i+k}^{\text{r}}}\right\} =%
\frac{J_{\text{MGP}}\left[ 1-\sum_{k=1}^{\bar{\ell}}\rho _{i+k}^{\text{r}%
}\right] }{\gamma _i\left[ 1-\sum_{k=1}^\ell \rho _{i+k}^{\text{r}}\right] }
\label{RR}
\end{equation}
and obtain the rest of the densities ($i=1,...,N-\ell $). Finally, to fix
the unknown $J_{MGP}$, we impose Eqn.~(\ref{exactJ1}): 
\begin{eqnarray}
J_{\text{MGP}}=\alpha \left\{ 1-\sum_{k=1}^\ell \rho _k^{\text{r}}\right\}
\label{end}
\end{eqnarray}
Though $J_{\text{MGP}}$ is ``just the solution to a polynomial equation,''
its exact value is quite intractable, since the order of the polynomial
approaches $2^N$ for large $\ell $ (such as $\ell =12$). Unfortunately,
numerical techniques are also of limited value, due to the extreme
sensitivity of the BR to small inaccuracies. As a result, given the
computational power of four decades ago, MGP were able to exploit this
approach only for a limited range of $\ell $ and $N$. Our interest here is
the edge effect associated with just one slow site near the entrance. So, we
would be considering a short sublattice ($L$, length $k\lesssim 30$) coupled
to a longer one ($R$). Thus, we are in an ideal position to exploit MGP's
approach for the $L$ sublattice, while matching it to the results of an
infinite TASEP for the $R$ sublattice.

Noting that the particle densities are uniform after the slow site (i.e., $\rho _{i>k}^{\text{r}} = \rho _R^{\text{r}} = \rho _R/\ell $
), we approximate the $R$
sublattice as an infinite system, so that Eqn.~(\ref{infJ}) for a
homogeneous TASEP applies: 
\begin{eqnarray}
J_R\simeq \frac{\rho _R(1-\rho _R)}{\ell -\bar{\ell}\rho _R}=\frac{\rho _R^{%
\text{r}}(1-\ell \rho _R^{\text{r}})}{1-\bar{\ell}\rho _R^{\text{r}}}
\label{J-rho2}
\end{eqnarray}
Of course, neither $J_R$ nor $\rho _R^{\text{r}}$ is known, and both must be
determined through matching conditions to the $L$ sublattice and $q$. For
the latter, we first consider expression~(\ref{MG_rec}) for site $k$: 
\[
J_{\text{MGP}}\simeq q\rho _k^{\text{r}}\frac{\rho _R^{\text{h}}}{\rho _R^{%
\text{r}}+\rho _R^{\text{h}}}=q\rho _k^{\text{r}}\frac{1-\ell \rho _R^{\text{%
r}}}{1-\bar{\ell}\rho _R^{\text{r}}} 
\]
where we have inserted the uniform density noted at sites beyond $k$. Of
course this must be $J_R$, which provides us with the reader density at the
slow site: 
\begin{equation}
\rho _k^{\text{r}}=\rho _R^{\text{r}}\,/q\,.  \label{rho-k}
\end{equation}
Continuing with the BR, we have 
\begin{eqnarray}
\rho _{k-i}^{\text{r}} &=&J_{MGP}\left\{ 1+\frac{\rho _{k-i+\ell }^{\text{r}}%
}{1-\sum_{n=1}^\ell \rho _{k-i+n}^{\text{r}}}\right\} \\
&=&\frac{\left( 1-\bar{\ell}\rho _R^{\text{r}}\right) \left[ 1-\sum_{n=1}^{%
\bar{\ell}}\rho _{k-i+n}^{\text{r}}\right] }{\rho _R^{\text{r}}\left( 1-\ell
\rho _R^{\text{r}}\right) \left[ 1-\sum_{n=1}^\ell \rho _{k-i+n}^{\text{r}%
}\right] }
\end{eqnarray}
until we have all the reader densities, $\rho _1^{\text{r}},...,\rho _k^{%
\text{r}}$, as functions of $\rho _R^{\text{r}}$. Finally, we impose Eqn.~(%
\ref{end}) for our case 
\begin{eqnarray}
1=\frac 1\alpha =\frac{\rho _R^{\text{r}}\left( 1-\ell \rho _R^{\text{r}%
}\right) }{1-\bar{\ell}\rho _R^{\text{r}}}\left[ 1-\sum_{n=1}^\ell \rho _n^{%
\text{r}}\right]  \label{end1}
\end{eqnarray}
which fixes the unknown $\rho _R^{\text{r}}$ and so, all quantities of
interest.

\begin{figure}[th]
\begin{center}
\hspace{-0.5cm} \includegraphics[height=6cm,width=8cm]{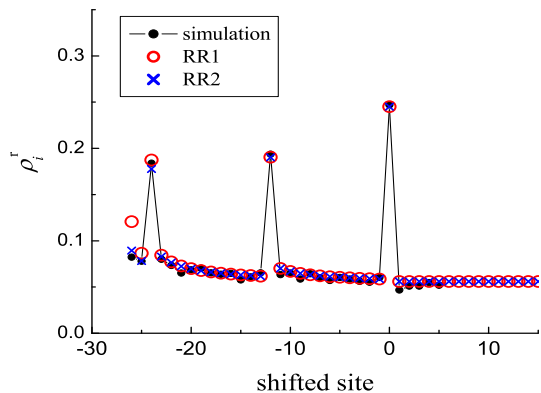}
\par
\end{center}
\caption{Density profile obtained through a BR relation. $q=0.2,k=26$ and $%
\ell =12$. The parameters used for fitting are $J$ and $\rho _R$. }
\label{fig:MG_rho}
\end{figure}

As an illustration, we carry out this program for the specific case of $%
\ell =12$, $k=26$ and $q=0.2$, and find \cite{ETD} 
\begin{equation}
\rho _R^{\text{r}}\cong 0.0506;\quad J_{\text{MGP}}\cong 0.0448\,\,.
\end{equation}
These values agree reasonably well with the simulation results of (0.0544,
0.0472), respectively. The density profile $\rho _i^{\text{r}}$ associated
with this result, shown in Fig.~\ref{fig:MG_rho}, is labled RR1 and by the
circles (red online). Apart from the first few sites, the fit is quite
respectable. However, the fit improves considerably if we \emph{arbitrarily }%
relax the constraint (\ref{end1}). In the same figure, we also display such
an alternative ($\times $'s, blue online, labled RR2), obtained by choosing 
\begin{equation}
\tilde{\rho _R^{\text{r}}}=0.0560\quad \Rightarrow \quad \tilde{J}_{\text{MGP%
}}=0.0478  \label{RR2}
\end{equation}
We found the $R^2$ coefficient of this fit improves from 0.97 to 0.99.
Moreover, it provides much better agreements with the $R$-sublattice density
as well as the overall current. The price, however, is a rather poor $%
\alpha $, with the right hand side of Eqn. (\ref{end1}) missing unity by $%
33\%$! It is unclear why the BR displays this peculiarity, although we
should perhaps not expect much better agreements, given that some
correlations are ignored in this approach. 

To summarize, we present the results of the two BR's in Table \ref
{tab:percentage_rr}. If we impose the constraint (\ref{end1}) seriously, we
see that the overall fits are tolerable. On the other hand, if we relax this
constraint, then there is substantial improvement on all quantities \emph{%
except} ``$\alpha $''. Clearly, this method, however unsystematic, manages
to capture much of the details of the edge effect. Unfortunately, the BR
relation fails to produce the long tails in the reader profiles (e.g.,
Fig.~5 of \cite{DSZ_PRE}). Indeed, Eqn.(\ref{MG_rec}) becomes very unstable
after about 40 steps, setting a limit on the maximum $k$ to which it can be
applied. We believe there are inherent difficulties with this approach
(beyond that of machine accuracy), but that is outside the scope of this
paper.
\begin{table}[tbp]
\caption{Summary of fit parameters and $\alpha $ for the two recursion
relation schemes. See text for the details. The percentage deviation from
the simulation is included in the parenthesis. The last row shows the value
of $R^2$ from the fit to the profiles in Fig. 2.}
\label{tab:percentage_rr}
\begin{center}
\begin{tabular}{|c|ccc|}
\hline
 & simulation & RR1 (\%) & RR2 (\%) \\ \hline
$\rho _R^{\text{r}}$ & 0.0550 & 0.0506 (8.00) & 0.0560 (-1.82) \\ 
$J$ & 0.0472 & 0.0448 (5.08) & 0.0478 (-1.23) \\ 
$\alpha $ & 1.0 & 0.95175 (4.82) & 1.5045 (50.45) \\ 
$R^2 $ &  & 0.97 & 0.99 \\ \hline
\end{tabular}
\end{center}
\end{table}

\subsection{\label{sub:fsmf}{Finite-segment mean-field (FSMF) theory}}

Given that we are interested in the edge effect, we can improve on the above
method by accounting for the physics of the small $L$ sublattice exactly.
This approach follows the work of Chou and Lakatos \cite{TomChou}, in which
the ``finite-segment mean-field theory'' was developed to understand \textit{%
quantitatively} the effects of clustered defects. Here, we generalize this
method to particles of size $\ell >1$ and solve the full master equation
explicitly for the $L$ sublattice (for small $k$). The key idea is to find
the exact expression for the current for this small \textit{finite} segment
and then match it to the result of an \textit{infinite} system (i.e., the $R$
sublattice). The approximations appear only in the matching conditions and
finite size, $O\left( 1/N\right) $, effects associated with the latter. We
further consider the interplay between the defect rate $q$ and the on-rate $%
\alpha $. Although our results are based on this simplified model,
understanding such interactions elucidates the effects of having a slow
codon or a cluster of slow codons near the initiation codon of an mRNA,
which is frequently observed in living organisms \cite{Chou11,Chou14}.

For the $L$ sublattice, the maximum dimension of the transition matrix is $%
2^{(k+\ell )}$, which requires enormous amounts of computing time even for $%
k=10$ and $\ell \leq 12$. Fortunately, for $k<\ell $, there can only be $%
\frac 12(k^2+3k+2\ell +2)$ allowed states in the $k+\ell $ segment. In
particular, for $k=1$, the dimension of the transition matrix is $\ell +3$
which is sufficiently small for us to understand the ``edge effect.'' We
illustrate this method through a detailed account for the simplest case ($k=1
$), showing the results for both $\ell =1$ and $\ell >1$ (as there are
subtle differences for the latter).

\begin{figure}[tbp]
\begin{center}
\includegraphics[height=3cm,width=10cm]{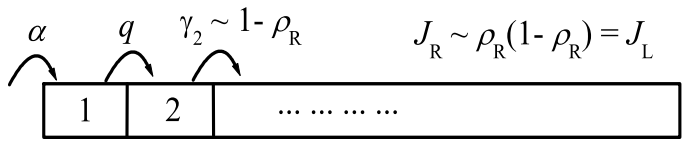} \vspace{-0.6cm}
\end{center}
\caption{Sketch of one slow site $q$ at $k=1$. FSMF matches a two-site TASEP
with the rest of the system. }
\label{fig:k=0}
\end{figure}

For the case of $k=\ell =1$, we consider \emph{two} sites, i.e., the entire $%
L$ sublattice and one site for the $R$ sublattice. Thus, we have only two
rates: the entrance rate $\alpha $ and the rate for the slow site $q$. There
are four possible configurations ($n_1=0,1;n_2=0,1$), labeled by $x_{0,1,2,3}
$. For convenience, we list these in their binary sequence, namely $x_0$
corresponds to state ($0,0$), $x_1$ to ($0,1$), etc. The master equation for
the evolution of $P_i$, the probability to find the system in $x_i$ at time $%
t$ is 
\begin{equation}
\partial _tP_i(t)=\sum_j[w_i^jP_j(t)-w_j^iP_i(t)]
\end{equation}
where $w_j^i$ is the rate of $x_i$ to $x_j$. Referring to Fig.~\ref{fig:k=0}
and writing the right hand side of the above as a matrix $\mathbb{W}$
operating on a vector $\vec{P}(t)$, we write explicitly 
\[
\mathbb{W}=\left( 
\begin{array}{cccc}
-\alpha  & \beta _L & 0 & 0 \\ 
0 & -\alpha -\beta _L & q & 0 \\ 
\alpha  & 0 & -q & \beta _L \\ 
0 & \alpha  & 0 & -\beta _L
\end{array}
\right) \,\,.
\]
Here, $\beta _L$ is an effective exit rate, which is to be fixed by
matching. The stationary state distribution $\vec{P}^{*}$ is easily found: 
\[
\vec{P}^{*}=Z^{-1}\left( 
\begin{array}{c}
\beta _L/\alpha  \\ 
1 \\ 
(\alpha +\beta _L)/q \\ 
\alpha /\beta _L
\end{array}
\right) 
\]
where $Z=1+\beta _L/\alpha +\alpha /\beta _L+(\alpha +\beta _L)/q$ is the
normalization factor. The steady state current $J_L$ follows readily: 
\begin{eqnarray}
J_L &=&\alpha \left\langle 1-n_1\right\rangle =q\left\langle n_1\left(
1-n_2\right) \right\rangle =\beta _L\left\langle n_2\right\rangle 
\label{JL1} \\
&=&(\alpha +\beta _L)/Z  \label{JL}
\end{eqnarray}
Meanwhile, on the $R$ sublattice, we have 
\[
J_R=\left\langle n_2\left( 1-n_3\right) \right\rangle =\left\langle
n_3\left( 1-n_4\right) \right\rangle =...
\]
which becomes $J_R=\rho _R(1-\rho _R)$ for an infinite system in the low
density phase. Matching this to the last equation in (\ref{JL1}), we arrive
at 
\[
\beta _L=1-\rho _R
\]
and so, 
\[
J_L=J_R=\beta _L\left( 1-\beta _L\right) \,\,.
\]
Setting this equal to expression (\ref{JL}), we find an equation for $\beta
_L$. The final answer for the current in this approximation scheme is 
\begin{equation}
J_{\text{FSMF}}\left( \alpha ,q;\ell =1\right) =\frac \alpha {2(q+\alpha
)}\{(1+\alpha )(q-\alpha )+\mathcal{R}\}  \label{Jaq}
\end{equation}
where $\mathcal{R\equiv }\sqrt{(1+\alpha )(q+\alpha )(\alpha ^2+\alpha
+q-3\alpha q)}$.

We caution that this formula should be used with some care. Though not very
transparent, this function monotonically increases with both $q$ and $\alpha 
$ when both are small. However, beyond a line in the $\alpha $-$q$ unit
square, it decreases back to zero. The maximum $J$ on this line is precisely 
$0.25$, at which point the system enters the MC phase. Substituting $0.25$
into the left of Eqn. (\ref{Jaq}), we find the phase boundary: 
\begin{equation}
q_c\left( \alpha \right) =\frac{\alpha (2\alpha +1)}{4\alpha ^2+2\alpha -1}
\label{trans}
\end{equation}
beyond which ($q\geq q_c$) the MC state prevails.

\begin{figure}[tbp]
\begin{center}
\includegraphics[height=6cm,width=8cm]{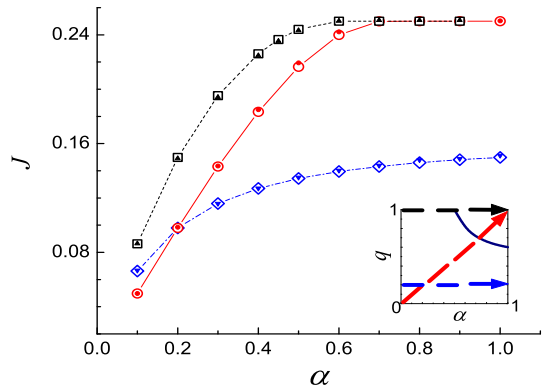} \vspace{-0.6cm}
\end{center}
\caption{Comparison between the simulations(solid symbols) and the FSMF
approximations from Eq.~\ref{Jaq} (connected open symbols). Top (black
online; $\blacktriangle $, $\Box $): $\alpha =1$; Middle (red online; $%
\bullet $, $\bigcirc $): $\alpha =q$; and Bottom (blue online; $%
\blacktriangledown $, $\diamond $): $q=0.2$. In all cases, $\ell =1$ and $%
N=1000$. The inset follows the same color coding scheme. See text for
details.}
\label{fig:ST}
\end{figure}

To further appreciate the quality of this theory, we present the comparison
between its predictions and simulation data in Fig.~\ref{fig:ST}.
Specifically, we show three typical scans through the $\alpha $-$q$ plane,
sketched in the inset of Fig.~\ref{fig:ST}. The upper-right corner
represents the MC phase (where $J=0.25$), with the solid (black online)
curve being $q_c\left( \alpha \right) $ in Eqn.~(\ref{trans}). The
color-coding in the inset matches the data in the main plot of Fig.~\ref
{fig:ST}. Here, the statistical error associated with the simulations is
estimated to be around $0.01\%$. We observe excellent agreement (within $1\%$%
) between the data and the theory results. Needless to say, we can extend
this approach, with some labor, to $k>1$. Since we doubt that the
agreement would be significantly different, we believe there is no need to
pursue this investigation further. Instead, we turn next to the more
interesting systems with extended objects.

Generalizing to the case with $\ell >1$, we need to account for two novel
aspects. One important difference is the current matching condition: Instead
of $J_L=\beta _L\simeq (1-\rho _R)$, we use the MGP result for $\ell >1$
case and 
\begin{equation}
\beta _L\simeq \frac{1-\rho _R}{\rho _{i+\ell }^{\text{r}}+1-\rho _R}\simeq 
\frac{\ell (1-\rho _R)}{\ell -\bar{\ell}\rho _R}\,\,.  \label{betaL}
\end{equation}
The other new item is that, since we expect some ``period-$\ell $''
structure, we could consider all $k$ up to $\ell $ and still restrict
ourselves to having only one particle in the $L$ sublattice. Following the
spirit of our analysis above, we also wish to account for the effects (on
the $L$ sublattice) due to a particle which just moved into the $R$
sublattice. With at most two particles in our finite segment, the exclusion
``at a distance'' means that we need to study a system with $k+\ell $ sites.
Fortunately, the configuration space (for, say, the reader occupations $%
\left\{ r_i\right\} $) is still manageably small. Thus, there is just one
0-particle state, $(k+\ell )$ 1-particle states, and $k\left( k+1\right) /2$
2-particle states. We demonstrate the case for $k=1$ here. Let us label the
0-particle state by $x_0$, the state with a reader in site $i$ by $x_{\ell
+2-i}$ ($i=1,...,\ell +1$), and finally, the 2-particle state by $x_{\ell
+2} $. To be pedantic, we show the explicit set of occupation numbers $%
\left\{ r_j\right\} $ corresponding to these $x$'s: 
\begin{eqnarray}
x_0 &\Leftrightarrow &\left( 
\begin{array}{cccccc}
0 & 0 & \cdots & \cdots & 0 & 0
\end{array}
\right)  \nonumber \\
x_1 &\Leftrightarrow &\left( 
\begin{array}{cccccc}
0 & 0 & \cdots & \cdots & 0 & 1
\end{array}
\right)  \nonumber \\
x_2 &\Leftrightarrow &\left( 
\begin{array}{cccccc}
0 & 0 & \cdots & \cdots & 1 & 0
\end{array}
\right)  \nonumber \\
&&\vdots \\
x_\ell &\Leftrightarrow &\left( 
\begin{array}{cccccc}
0 & 1 & \cdots & \cdots & 0 & 0
\end{array}
\right)  \nonumber \\
x_{\ell +1} &\Leftrightarrow &\left( 
\begin{array}{cccccc}
1 & 0 & \cdots & \cdots & 0 & 0
\end{array}
\right)  \nonumber \\
x_{\ell +2} &\Leftrightarrow &\left( 
\begin{array}{cccccc}
1 & 0 & \cdots & \cdots & 0 & 1
\end{array}
\right)  \nonumber
\end{eqnarray}
The advantage of this slightly peculiar labeling is that it reduces to the $%
\ell =1$ case easily.

To find the current $J(\alpha ,q,\ell )$ for $\ell >1$, we need to compute
the new transition matrix $\mathbb{W}$. With the configurations clear in our
minds, we simply write: 
\begin{equation}
\mathbb{W}=\left( 
\begin{array}{cccccccc}
-\alpha & \beta _L & 0 & \cdots & \cdots & \cdots & 0 & 0 \\ 
0 & -\alpha -\beta _L & \beta _L & 0 & \cdots & \cdots & \vdots & \vdots \\ 
0 & 0 & -\beta _L & \beta _L & 0 & \cdots & \vdots & \vdots \\ 
\vdots & \vdots & 0 & -\beta _L & \beta _L & 0 & \vdots & \vdots \\ 
\vdots & \vdots & \vdots & \vdots & \vdots & \vdots & \vdots & \vdots \\ 
0 &  &  &  &  & -\beta _L & q & 0 \\ 
\alpha & 0 & \vdots & \vdots &  &  & -q & \beta _L \\ 
0 & \alpha & 0 & 0 &  &  & 0 & -\beta _L
\end{array}
\right) \,\,  \label{bigW}
\end{equation}
Similar to the $\ell =1$ case, we find 
\[
\vec{P}^{*}=Z^{-1}\left( 
\begin{array}{c}
(\beta _L/\alpha ) \\ 
1 \\ 
(\alpha +\beta _L)/\beta _L \\ 
\vdots \\ 
(\alpha +\beta _L)/\beta _L \\ 
(\alpha +\beta _L)/q \\ 
\alpha /\beta _L
\end{array}
\right) 
\]
to be the same, except for $\bar{\ell}$ more entries of $(\alpha +\beta
_L)/\beta _L$ in the middle. Thus, 
\[
Z=1+\beta _L/\alpha +\alpha /\beta _L+(\alpha +\beta _L)/q+\bar{\ell}(\alpha
+\beta _L)/\beta _L 
\]
Meanwhile, we still have $J_L=(\alpha +\beta _L)/Z$. Finally, matching $J_L$
with $J_R=\rho _R(1-\rho _R)/\left( \ell -\bar{\ell}\rho _R\right) $ and
using Eqn.~(\ref{betaL}), we arrive at the solution for general $\ell $ (and 
$k=1$ ): 
\begin{equation}
 J_{\text{FSMF}}(\alpha ,q,\ell )=\frac \alpha {2(q+\alpha )}\frac{q-\alpha +%
\mathcal{R}(1+\alpha )^{-1}+\bar{\ell}\left[ q(1-\alpha )+\mathcal{R}q\left(
q+\alpha \right) ^{-1}\right] }{(1+\alpha )^{-1}+\bar{\ell}\left[ 1+\bar{%
\ell}\alpha q(q+\alpha )^{-1}\right] }  \label{J_ell_k.....}
\end{equation}
We have written $J_{\text{FSMF}}$ in a form that clearly reduces to Eqn. (%
\ref{Jaq}) for $\ell =1$. Given the shifted phase boundaries for $\ell >1$,
the system enters the MC phase when: 
\begin{equation}
q_c=\frac{\alpha \hat{\chi}(\alpha +1-\hat{\chi})}{\alpha (\alpha +1-\hat{%
\chi})-\hat{\chi}(1-\hat{\chi})}
\end{equation}
In the next section, we will summarize our findings along with some
comparisons to Monte Carlo simulation data.

\section{\label{sum}Summary and outlook}

We investigated how a single defect site near the lattice boundary (small $k$%
) influences the steady state properties of the system for both point
particles and those of length $\ell >1$. The simplest ``mean-field''
approaches -- NMF and SKL -- are unsuitable, since both rely on matching two
infinite TASEP's across a defect and cannot address the issue of $k$%
-dependence. Instead, we considered two more sophisticated levels of
mean-field methods, with complementary strengths and weaknesses. One method,
first used by MacDonald, et. al. \cite{MG} (MGP), is based on a recursion
relation for the density profile. The advantage of MGP is that, up to
moderate $k$ values, its predictions for both the profile and the current
are reasonably good. The weakness is that we can access these predictions
only numerically so that the dependence on the control parameters $\alpha
,q,k,$ and $\ell $ remains obscure. It is also unclear how to systematically
improve on this approach. The other method, based on an exact account of the
physics of the first $k+\ell $ sites, is a generalization of the finite
segment mean-field (FSMF) theory of Chou and Lakatos \cite{TomChou}. The
strengths of this method are many. Based on the steady state solution to the
full master equation, it can be improved systematically. Its predictions
agree with simulations exceedingly well and provide analytic expressions so
that the dependence on $\left( \alpha ,q,\ell \right) $ can be appreciated.
Both methods are obviously severely restricted to a relatively small range
of $k$'s. In MGP, the limitation arises from the extreme sensitivity of the
recursion relation and, in FSMF, an exponential increase (in the worst
scenario) in the size of the transition matrix. Moreover, some of the long
tails in the ``edge effect'' extend up to $k\sim 50$ \cite{DSZ,DSZ_PRE},
well beyond the present reach of either approach. Hopefully, more efficient
approaches will be developed in the future.

To summarize, we find three successively better methods to describe a TASEP
with a defect near the entrance, for $\ell \geq 1$. To illustrate, we
present the results of all three mean-field approaches, along with
simulation data, in Tables \ref{tab:ell1} and \ref{tab:ell2}. Both concern
the case with $\alpha =\beta =1$ and $k=1$; the difference being $\ell =1,2$
in the two Tables. We see that, by accounting for exclusion at a distance,
MGP (with the fit through $RR1$) clearly succeeds better than NMF when $\ell >1$. Meanwhile, it is hardly
surprising that an \emph{exact} treatment of the finite segment before the
defect is superior to both. 

Though not displayed explicitly, similar
improvements are found to hold for $\ell $ up to $12$. Finally, if we use $%
J_{\text{FSMF}}$ for $k=1$ and $J_{\text{NMF}}$ for $k\gg 1$ (i.e., deep in
the bulk), we arrive at a prediction for $\Delta _1\left( q\right) \equiv
\left. J_q(k=1)\right/ J_q(k\rightarrow \infty )$, defined in \cite{DSZ_PRE}%
. The remarkable non-monotonic behavior in $\Delta _1$ (shown in Fig. 7 of
\cite{DSZ_PRE}) is well captured by the
combination of these two mean-field approaches.

\begin{table}[tbp]
\caption{Different mean-field approximations for the current $J(q)$ for the
case $\alpha =\beta =1$, $k=1$ and $\ell =1$. Simulation results are based
on a lattice with $L=1000$.}
\label{tab:ell1}
\begin{center}
$
\begin{tabular}{|c|cccc|}
\hline
$q$ & $J_{\text{NMF}}$ & $J_{\text{MGP}}$ & $J_{\text{FSMF}}$ & simulation
\\ \hline
0.1 & 0.0826 & 0.0833 & 0.0863 & 0.0864 \\ 
0.2 & 0.1389 & 0.1421 & 0.1498 & 0.1490 \\ 
0.3 & 0.1775 & 0.1848 & 0.1954 & 0.1941 \\ 
0.4 & 0.2041 & 0.2153 & 0.2261 & 0.2248 \\ 
0.5 & 0.2222 & 0.2361 & 0.2440 & 0.2432 \\ 
0.6 & 0.2344 & 0.2476 & 0.2500 & 0.2502 \\ \hline
\end{tabular}
$
\end{center}
\end{table}
\begin{table}[tbp]
\caption{Different mean-field approximations for the current $J(q)$ for the
case $\alpha =\beta =1$, $k=1$, and $\ell =2$. Simulation results are based
on a lattice with $L=1000$.}
\label{tab:ell2}
\begin{center}
\begin{tabular}{|c|cccc|}
\hline
$q$ & $J_{\text{NMF}}$ & $J_{\text{MGP}}$ & $J_{\text{FSMF}}$ & simulation
\\ \hline
0.1 & 0.0758 & 0.0763 & 0.0788 & 0.0771 \\ 
0.2 & 0.1190 & 0.1213 & 0.1266 & 0.1231 \\ 
0.3 & 0.1442 & 0.1484 & 0.1543 & 0.1509 \\ 
0.4 & 0.1587 & 0.1639 & 0.1680 & 0.1665 \\ 
0.5 & 0.1667 & 0.1708 & 0.1715 & 0.1717 \\ \hline
\end{tabular}
\end{center}
\end{table}


Beyond our investigations here, there is ample room for future research. In
an open TASEP, there are two ``edges'' and so, two possible ``edge
effects.'' We reported findings for only one. When the slow site is near the
exit ($k\simeq L$), the current is also observed to increase \cite{ETD}.
However, due to lack of particle-hole symmetry for $\ell >1$, this increase
is not the same as the case for small $k$. Further, there are serious
complications associated with the profile, especially for small $q$ (e.g.,
Fig. 5 in \cite{DSZ_PRE}). Thus, it would be desirable to find better
methods to understand these peculiarities quantitatively. Similarly, we
should explore the ``edge effect'' for $q>1$. When a ``fast site'' is deep
in the bulk, it has little effect on the current. However, its effects if
located near the edges, especially near the exit end, are yet to be
discovered. Beyond one defect, there are obvious questions concerning two or
more defects. It was found that two equally slow sites deep in the bulk
``interact'' \cite{DSZ_PRE}, in the sense that the overall current is
significantly suppressed when they are located near each other. Much less
has been investigated when the two defects are associated with \emph{%
different} rates. Both mean-field methods can easily be extended to study
such issues, especially when the two sites are near each other. At the other
extreme, we face a completely inhomogeneous TASEP. But this is precisely the
scenario more relevant for protein synthesis in vivo. In this sense, there
is much to be done before we reach the goal of a realistic model for
understanding the biological process of translation.

\section*{Acknowledgments}
We thank Andrea Apolloni, Rahul Kulkarni, Uwe T\"{a}uber, Brenda Winkel
for discussions, and especially Tom Chou and Rosemary Harris for
enlightening suggestions. One of us (RKPZ) thanks H.W. Diehl for his hospitality at Universit\"{a}t Duisburg-Essen 
and S. Dietrich at the Max Planck Institute fur Metallforschung, where some of this 
work was performed. This work is supported in part by the NSF through DMR-0414122, 
DMR-0705152, and DGE-0504196. JJD also acknowledges generous support from the 
Virginia Tech Graduate School.

\end{document}